\documentclass[twoside,leqno,twocolumn]{article}  
\usepackage{ltexpprt} 
\usepackage{graphicx}
\usepackage{amsmath}

\begin{document}

\title{\Large Event Discovery in Time Series}
\author{Dan Preston\thanks{Initiative in Innovative Computing, Harvard University} $^\ddagger$ \\
\and 
Pavlos Protopapas$^*$\thanks{Harvard-Smithsonian Center for Astrophysics}
\and
Carla Brodley\thanks{Department of Computer Science, Tufts University}}
\date{}

\maketitle

%\pagenumbering{arabic}
%\setcounter{page}{1}%Leave this line commented out.

\begin{abstract} \small\baselineskip=9pt The discovery of events in time series can have important implications, such as identifying microlensing events in astronomical surveys, or changes in a patient's electrocardiogram. Current methods for identifying events require a sliding window of a fixed size, which is not ideal for all applications and could overlook important events. In this work, we develop probability models for calculating the significance of an arbitrary-sized sliding window and use these probabilities to find areas of significance. Because a brute force search of all sliding windows and  all window sizes would be computationally intractable, we introduce a method for quickly approximating the results. We apply our method to over 100,000 astronomical time series from the MACHO survey, in which 56 different sections of the sky are considered, each with one or more known events. Our method was able to recover $100\%$ of these events in the top $1\%$ of the results, essentially pruning $99\%$ of the data. Interestingly, our method was able to identify events that do not pass traditional event discovery procedures. \end{abstract}

%The discovery of events in astronomical time series data is a non-trival problem. Existing methods address the problem by requiring a fixed-sized sliding window which, given the varying lengths of events and sampling rates, could overlook important events. In this work, we develop probability models for finding the significance of an arbitrary-sized sliding window, and use these probabilities to find areas of significance. In addition, we present our analyses of major surveys archived at the Time Series Center, part of the Initiative in Innovative Computing at Harvard University. We applied our method to the time series data in order to discover events such as microlensing in the MACHO, OGLE, and TAOS surveys. The analysis shows that the method is a useful tool for filtering out nearly 99\% of noisy and uninteresting time series from a large set of data, but still provides full recovery of all known microlensing and blue star events. Furthermore, due to its efficiency, this method can be performed on-the-fly and will be used to analyze upcoming surveys, such as Pan-STARRS.

%------------------------------------------------------------------------- 
\section{Introduction}

Event discovery in time series data is the focus of many modern temporal data mining methods \cite{anomaly_detection1, mining_deviants, hotsax}. An \emph{event} is characterized by an interval of measurements that differs significantly from some underlying baseline. The difficulty in identifying these events is that one must distinguish between events that could have occurred by chance and events that are statistically significant. In this paper, we present a method that is noise independent and determines the significance of an interval.

\begin{figure}[h] %  figure placement: here, top, bottom, or page
   \centering
   \includegraphics[width=3.25in,clip=true,trim = 0.8in 0.8in 0.43in 0.8in]{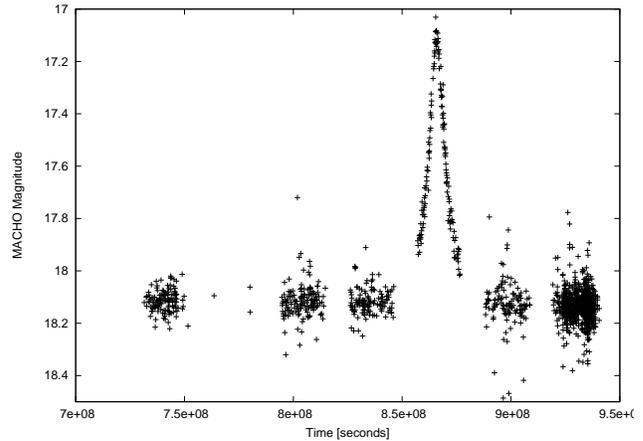}
   \caption{Example: A microlensing event in the MACHO data set, with ID 104.20121.1692. The X-axis represents time in seconds; the Y-axis represents magnitude.\protect\footnotemark[2]}
   \label{fig:event_example_ml}
\end{figure}

We focus our tests in astronomy, for which discoveries could be found by identifying events in time series data. For example, microlensing events occur when an object passes in front of a light source and acts as a gravitational lens, thus increasing the magnitude of light being observed. Figure \ref{fig:event_example_ml} is an example of such an event. Each light curve (time series data recorded by astronomical observations) has different noise characteristics, making it difficult to form a general noise model for all time series in the set.%
\footnote
   {%
     The reasons for varying noise characteristics in astronomical observations are plentiful: weather (clouds, wind, etc) create trends, the temperature can change the CCD characteristics, different stresses on the mechanical structure depending on the direction of the telescope can stress the optics differently and consequently make a difference in how the light is bent, and other possible effects of the environment.
   }
Furthermore, there are millions of observed light curves in modern astronomical surveys, many with uninteresting fluctuations or trends, thus any effective method must be able to distinguish between these variations and statistically significant events such as microlensing, flares and others. Examples of other applications for which event detection is useful include searching for events in stock market data \cite{discovery_stock1, discovery_stock2}, examining CPU usage to identify anomalies, and finding irregularities in medical data \cite{discovery_medical1, discovery_medical2}.

\footnotetext[2]{The magnitude of an observation is related to the logarithm of the intensity of the light being observed for each star.}

Searching for events in time series requires first searching for \emph{overdensities}, intervals of time series with a large deviation from the baseline \cite{spatial_clusters}, and then determining the statistical significance of the region. A na\"{i}ve approach would be to explore all possible intervals in order to identify the region with the largest deviation from the baseline. More efficient methods for identifying these regions have been explored in  \cite{bayesian_spatial_scan} and \cite{spatial_clusters}.

After determining regions of the time series that deviate from the baseline, one must determine if these regions are statistically significant, which requires modeling the noise of the time series.  Modern methods create such models by performing randomization testing, in which a time series is randomly reshuffled several times and the original interval is compared to the most anomalous interval found in each shuffle \cite{spatial_clusters}. The downfall of performing this randomization testing is that it requires significant computation, and needs to be performed for each new time series. This technique is intractable for domains in which we have thousands or even millions of time series, such as light curves in astronomy.

Our approach falls under the broader category of scan statistics \cite{scanstats, kulldorff}, which claims that by considering \emph{sliding windows}, or intervals of a time series, one can determine statistical significance if the underlying noise can be understood. To remove the need to model the noise for each time series, we begin by first converting the time series to \emph{rank-space}, a discrete series of $N$ points representing the rank of each value in a real-valued time series, where $1$ is the lowest value and $N$ is the highest. This creates a uniform distribution of points across the Y-axis, independent of the underlying noise. This allows for a probability model to be formed that does not require a model of the noise in the time series (this probability model is described in Section \ref{probability_dist}). Because the model is applicable to any time series of length $N$ that has been converted to \emph{rank-space}, it has the added benefit that it allows an analyst to compare the significance of events across different time series. 

Given the probability model described, each interval has a $p$-value, which represents the likelihood of the region occurring by chance. The method described in this paper considers all possible intervals of a time series. In other words, a variable window size is used in the analysis, and a $p$-value can be assigned to each window size. To make this search tractable, a well-known optimization technique \cite{powell} is applied that approximates the solutions, by finding the minimum $p$-value for all possible intervals.

In applying our method, we ran our method on the MACHO data set \cite{macho_original}. Out  of $110,568$ time series, our method was able to recover all known microlensing and blue star events \cite{macho_microlensing, macho_bluestar}. Furthermore, many of the events found in the top results would not have passed traditional tests, such as a $\chi^2$ test. Results of the MACHO analysis are explained further in Section \ref{macho_results}. In addition, Section \ref{synthetic_ts} details the analysis of $2000$ synthetic time series, in which all events were found, including zero false positives.

Finally, we compare the proposed approach to HOT SAX, an anomaly detection algorithm \cite{hotsax}. In doing so, we outline the essential differences between an anomaly detection algorithm (i.e., HOT SAX) and an event detection algorithm.

In the following sections, we begin by describing the related work in scan statistics and anomaly detection in Section \ref{related_work}. We then describe our method for forming a probability model to assess the statistical significance of an interval in Section \ref{assess_significance}, and how one can identify events based on these probability models. In Section \ref{optimization}, we describe how we can apply these methods to large datasets by approximating results. Finally, Section \ref{results} reports our results from an application to an existing astronomical survey, MACHO, and a study of synthetic time series. 

%------------------------------------------------------------------------- 
\section{Related Work}
\label{related_work}

Perhaps the work most closely related to event detection in time series is scan statistics, whose goal is to determine the probabilities and statistics of spatial subsets. Scan statistics are applied in most cases to temporal data, but in some cases are generalized to any number of $d$ dimensions. The goal of scan statistics is to discover statistically significant densities of points that deviate from some underlying baseline (see \cite{scanstats} for a detailed review). Scan statistics have been used to refute decisions made based on false identification of seemingly significant events. Examples of such events include the grounding of all F-14s by the U.S. Navy after a cluster of crashes, and when a judge fined a Tennessee State Penitentiary Institution after an unusual number of deaths occurred at the facility \cite{scanstats}. Scan statistics were used in both cases to show that the cluster in question was likely to have occurred by chance, and thus was not statistically significant.

To see how they are applied to time series data, consider a time series with $Z$ points, each independently drawn from the uniform distribution of time intervals on $[0,1)$ \cite{scanstats}. The \emph{scan statistic} $S_w$ is the largest number of points to be found in any interval of length $w$, thus $S_w$ is the \emph{maximum cluster} with window size $w$ \cite{kulldorff}. The determination of the distribution of the statistic $P(k | Z, w)$ is fundamental to this analysis, where $P(k | Z, w)$ represents the probability of observing $k$ points with the window $w$. Much work has been done to find approximate results for this statistic, such as the use of binomial probabilities \cite{wallenstein_scanstats}, and Poisson distributions \cite{scanstats_poisson_dist}. \nocite{glaz_scanstats_dist} Generally, these approximations work well for small values of $w$ and $k$, with the varying levels of complexity and accuracy. It is important to note that in many applications, a rough approximation is sufficient to find significance. Using this statistic, one can determine if an interval of the time series is not consistent with the noise, and thus is statistically significant. In this work, we describe a new statistic for assessing the statistical significance of any arbitrary window by forming a new probability distribution, motivated by work in scan statistics, that will scale to large values of $w$. Our method has an analytic method for finding the exact probability distribution. Finally, by forming the distribution for all window sizes, one can analyze significance for any arbitrary window size.

Existing methods in scan statistics currently do not address the problem of large data sets (e.g., millions or billions of time series). In many applications, one requires an event detection algorithm that can be performed quickly on many time series. In addition, because much of the data for these applications has noise that is difficult to understand and model, it is critical to develop a method that is independent of the noise characteristics. The current methods in scan statistics are generally based on particular noise models, such as Poisson distributions \cite{scanstats_poisson_dist} and binomial probabilities \cite{wallenstein_scanstats}. Both of these characteristics exist in astronomy, and thus our goal in this paper is to address both of these issues. Due to the intractability of current scan statistics methods, we cannot compare our method. Thus, we are compelled to compare to anomaly detection due to its speed, efficiency and that it does not require first modeling the noise.

Anomaly detection in time series is closely related to the problem of searching for events in time series. In \cite{hotsax}, time series discords are used to determine the interval most likely to be an anomaly. Keogh et al present a method for determining these discords for a fixed-size sliding window. At its core, the method considers all possible intervals of the time series for a given size (a parameter provided by the domain expert), and determines the subset with the largest distance from its nearest \emph{non-self match} (a distinction made to avoid trivial matches). Requiring the analyst to provide only one parameter is a significant improvement over past methods which require several unintuitive parameters \cite{anomaly_detection2}. Defining a window size in advance is realistic when the expert has knowledge about the size of the event, or when the time series is periodic and the event is likely to occur during one period. As an example, this is the case when analyzing an electrocardiogram time series \cite{hotsax}. On the other hand, when an expert does not know the characteristics of the event or when the event could occur at different intervals, it is preferable to examine any arbitrary window size. Such is the case in astronomy, where one may be searching for any novel discovery represented by possible significant changes in light patterns (e.g., microlensing) that each may have a different time interval.

It is important to understand the distinction between anomaly detection and event detection. When detecting anomalies, such as in \cite{hotsax}, the anomalies are discords from the time series (or set of time series). This is fundamentally different from the goal of this paper, which is to find subintervals that are statistically significant from the underlying noise. In other words, anomaly detection finds those subintervals that differ most from the rest of the ensemble, whereas event detection searches for those that significantly differ from the noise.

Methods of identifying events in astronomical data exist in the literature. Traditionally for fast identification of variables, a $\chi^2$ test of a straight line has been used, but overlooks events of short duration. Other methods involve fitting event shapes to the light curves, generally using a least-squared fit \cite{finding_variables}. This can be thought of as creating synthetic light curves with the characteristics of known events, and testing these hypotheses. The fitting of events is an accurate method for event discovery, yet it is slow and quite expensive. What makes our method particularly valuable is that it can find events that escape the $\chi^2$ test and is less expensive than event-fitting tests. 

%Furthermore, we wish to consider $W_k$, which is the size of smallest window containing $k$ points. An important distinction to note is that the distributions of $S_w$ and $W_k$ are related such that $P(S_w \geq k) = P(W_k \leq w)$ \cite{kulldorff}. 

While scan statistics build a sound framework for determining the significance of events, there are significant challenges that must be overcome in order to deal with large amounts of high-dimensional data. Our work addresses these issues by first performing a transformation of the data to a uniform representation and second, developing a probability distribution for assessing the significance of intervals of a time series. Furthermore, although anomaly detection is related to event detection, it is fundamentally different and not particularly well suited for our application. Our empirical results in Section \ref{macho_results} make the distinction between anomaly and event detection clear. 

%------------------------------------------------------------------------- 
\section{Assessing the Statistical Significance of an Interval}
\label{assess_significance}

The following work differs from the work in scan statistics in three key ways. First, we do not consider the temporal distribution of individual events, but instead \emph{bin} the events in equidistant time intervals. In other words, a time series with $Z$ points distributed on $[0,1)$ as defined for scan statistics in Section \ref{related_work} is transformed into a time series by grouping the points into $N$ bins, with each bin $b$ representing a constant interval of time. Each bin becomes a value $v_b$ in the time series, where $v_b$ is the number of points for the bin $b$. Second, we are not concerned in particular with the largest subset $S_w$, but with any interval with statistical significance. Third, we consider any arbitrary window size $w$, instead of using a fixed value. Forming a distribution for all window sizes will allow an analyst to identify the exact interval in question. More importantly, this step introduces a method for the comparison of $p$-values for any arbitrary window size. In this section, we present a method for simplifying our problem by converting a time series into \emph{rank-space}, and then two methods for creating a probability distribution of possible window sums for assessing the significance of an interval.

\subsection{Formal Definition.}

%	\begin{equation}
%	T = \left[\begin{array}{c} v_1, v_2, \ldots, v_t, \ldots, v_N \end{array} \right]
%	\end{equation}

%\begin{equation}
%S = \left[\begin{array}{c} v_s, v_{s+1}, \ldots, v_{s+w-1} \end{array} \right]
%\end{equation}

We begin by defining a time series $T = [v_1, v_2, \ldots, v_t, \ldots, v_N]$ of length $N$ where $v_t$ is the value of a time series $T$ at time $t$. It is assumed that the length of each time interval $[t, t+1]$ is equal for all $1 \leq t < N$. An interval $S \subseteq T$ is defined by a starting point $s$ and a window size $w$. Formally, $S = [v_S, v_{s+1}, \ldots, v_{s+w-1}]$ where $1 \leq s \leq s+w-1 \leq N$.

\subsection{Rank-Space Transformation.}
\label{rankspace}

In order to remove the need to model the noise for each individual time series, one can consider a time series $T_R$ that has been transformed into \emph{rank-space}. To this end, we rank each point in a time series $T$ from $1$ to the number of total points $N$, where the lowest value is assigned a value of $1$ and the largest is assigned a value of $N$. This will allow us to consider a time series with a uniform distribution of noise, no matter what the noise distribution was before the transformation.

%	\begin{equation}
%	Q(s,w) = r_s + r_{s+1} + \cdots + r_{s+w-1}
%	\end{equation}

By examining the sums of the values inside any arbitrary sliding window of $T_R$, we can determine its $p$-value (the probability that the sum occurred by chance). How to calculate the $p$-value is the focus of Sections \ref{probability_dist} and \ref{probability_dist2}. Consider a starting point $s$ and a window size $w$, and let $Q(s,w)$ be the sum of the ranked values inside that window, where $1 \leq s \leq N$ and $1 \leq s+w-1 \leq N$. Thus, we define the sum $Q(s,w) = r_s + r_{s+1} + \cdots + r_{s+w-1}$ where $r_s, \ldots, r_{s+w-1} \in T_R$.
Our method is based on the assumption that the sums in the outer tails of the probability distribution of all possible sums are significant, as they deviate drastically from the underlying baseline.

Note that with \emph{rank-space} one loses the depth of the time series. In other words, the amount by which an event deviates from the baseline is no longer distinguishable after the transformation, due to the uniform spacing of the points. For example, a time series with a significant event that deviates from the baseline with a signal to noise ratio of $5$ is no different than an event with a signal to noise ratio of $20$ in \emph{rank-space}. This is beneficial in many cases in which only a few points deviate by large amounts from the baseline, but are not necessarily significant events.

A distinct probability distribution exists for each pair of $(w,N)$, because each sum is dependent only on the number of values being added together ($w$), and their bounds, $1 \leq r_i \leq N$. Therefore, in order to assess the statistical significance of a particular sum $Q(s,w)$ of $T_R$, we must know the distribution of $(w, N)$. We first describe an analytic method to find this distribution, and due to its computational complexity, a Monte Carlo method for approximating the same distribution. 

%------------------------------------------------------------------------- 
\subsection{Determining the Probability Distribution: Analytic Method.}
\label{probability_dist}
% in integers: $\in \mathbb{Z}$

We wish to find the probability that any sum $\phi$, obtained by the performing the \emph{rank-space} transformation described above, would appear in uniformly random noise with a window size $w$ and a maximum value $N$. Thus, our goal is to produce an exact probability distribution function for a given $\phi$, $w$ and $N$.

To obtain our probability curve, we can count the number of ways any $\phi$ can be formed with distinct values from $1, \ldots, N$ inclusive, and divide by ${N \choose w}$ (all combinations of $w$ distinct values, which represents all ways to form $\phi$ with $w$ values).

To find the number of ways a single sum $\phi$ can be formed, we consider the combinatorial problem of finding the number of partitions of $\phi$ with $w$ distinct parts, each part between $1$ and $N$, inclusive.  In other words, we wish to count all possible solutions to $v_1 + v_2 + \cdots + v_w = \phi$ where $0 < v_1 < v_2 < \ldots < v_w \leq N$.

In order to solve this, we transform the problem to match a well-known problem involving a specific application of the $q$-binomial coefficients \cite{theory_of_partitions}. This requires that we slightly modify the problem. The inequality above is the same as the following: First, subtract 1 from the smallest part, 2 from the second, etc. to get $0 \leq v_1 - 1 \leq v_2 - 2 \leq ... \leq v_w - w \leq N-w$. We have subtracted a total of $1 + 2 + \ldots + w = w(w+1)/2$, so we are now looking for the number of partitions of $\phi - w(w+1)/2$ with at most $w$ parts, and with largest part at most $N-w$.            

Following the notation of \cite{theory_of_partitions}, we define $p(n, m, k)$ to be the number of partitions of $k$ into at most $m$ parts, each part at most $n$. This value is the coefficient of $q^k$ in the generating function,
\begin{equation}
	G(n, m; q) = \sum_{k \geq 0} p(n, m, k)\,q^k
\end{equation}
\noindent where

\begin{equation}
             G(n, m; q) = \frac{(1-q^{n+m})(1-q^{n+m-1}) \cdots (1-q^{m+1})}{(1-q^n)(1-q^{n-1})\cdots(1-q)}
\end{equation}
%\begin{eqnarray}
%            G(n, m; q) & = &  \\
%            & \frac{(1-q^{n+m})(1-q^{n+m-1}) \cdots (1-q^{m+1})}{(1-q^n)(1-q^{n-1})\cdots(1-q)} & \nonumber
%\end{eqnarray} % \nonumber

	  A basic identity of the coefficients provides the recurrence on page 34 of \cite{theory_of_partitions}:
\begin{equation}
	p(n, m, k) = p(n-1, m, k-m) + p(n, m-1, k)
\end{equation}
\noindent where the base case follows from the identity

\begin{equation}
	p(n, 0, k) = p(0, m, k) = 
\left\{
\begin{array}{lr}
1 & \mathrm{if\ } n = m = k = 0 \\
0 & \mathrm{otherwise}
\end{array}
\right.
\end{equation}
\noindent Applying this to our problem, we are looking for distinct partitions of $k = \phi-w(w+1)/2$ with at most $m = w$ parts and largest part at most $n = N-w$. Thus, $p(N-w, w, \phi-w(w+1)/2)$ is the number of ways to create the sum $\phi$ with $w$ unique parts, each value from $1$ to $N$. The probability of obtaining sum $\phi$ will equal:
\begin{equation}
	P(\phi; w, N) = \frac{p(N-w, w, \phi-w(w+1)/2)}{{N \choose w}}
\end{equation}
Thus, to build a distribution for $(w,N)$, we find $P(\phi; w, N)$ for all $\phi$. %$\frac{w(w+1)}{2} \leq \phi \leq Nw - \frac{w(w-1)}{2}$

%------------------------------------------------------------------------- 
\subsection{Determining the Probability Distribution: Monte Carlo Method.}
\label{probability_dist2}

Although the analytic results are preferable and provide exact probabilities for all possible sums, the analytic method is deeply recursive and prohibitively expensive. Memoization can be used to improve performance, by creating a hash for pruning branches of the recursion tree, but the hash still remains too large in memory to be practical in all cases. In order to perform the memoization method for creating the analytic distribution, one must keep a three-dimensional structure in memory to keep track of the recurrence of the size $w_{max} \times N \times \displaystyle \left( Nw_{max} - \frac{w_{max}(w_{max}-1)}{2} \right)$, where $w_{max}$ is the largest window size we wish to search for. For example, for $w_{max}=50$ and $N = 500$, we require 2.2GB of memory, which may be too large for some systems.

Thus an alternative and less memory demanding method is to perform random sampling on all possible sums. To perform the Monte Carlo method, we repeatedly choose $w$ unique random numbers from $1$ to $N$, sum them, and then count the frequency with which each sum appears. Dividing each of the frequency counts by the total number of samples gives an approximate probability distribution curve. 

Because the tails of the distribution represent the windows with the lowest $p$-values, those counts must be determined as accurately as possible. We seek statistically significant events, thus we define a threshold $\alpha$ for which we would like to consider all events with $p$-values lower than $\alpha$. Our goal in the accuracy of the probability distribution is to keep the $p$-values around the threshold $\alpha$ as accurate as possible. In other words, the accuracy of the $p$-value associated with the event is not imperative; it is ensuring that the $p$-values of events do not fluctuate around the threshold $\alpha$. In a Monte Carlo run, we consider $n_\phi$ to be the frequency for each sum $\phi$. The expected accuracy of the frequency of $\phi$ is given by
\begin{equation}
	\epsilon \sim \frac{1}{\sqrt{n_\phi}}
\label{eq:nQ}
\end{equation}
\noindent where $\epsilon$ is the error \cite{numerical_recipes}. In order to ensure accuracy of $\epsilon$ for a threshold $\alpha$, we must obtain a minimum of $\lambda$ samples for $\phi$, given by
\begin{equation}
	\lambda \sim \frac{n_\phi}{\alpha} \sim \frac{1}{\alpha \epsilon^2}
\end{equation}
\noindent because we know that $n_\phi = \frac{1}{\epsilon^2}$ from Equation \ref{eq:nQ}.
  
\begin{figure}[t] %  figure placement: here, top, bottom, or page
   \centering
   \includegraphics[width=3.25in,clip=true,trim = 0.8in 0.8in 0.43in 0.8in]{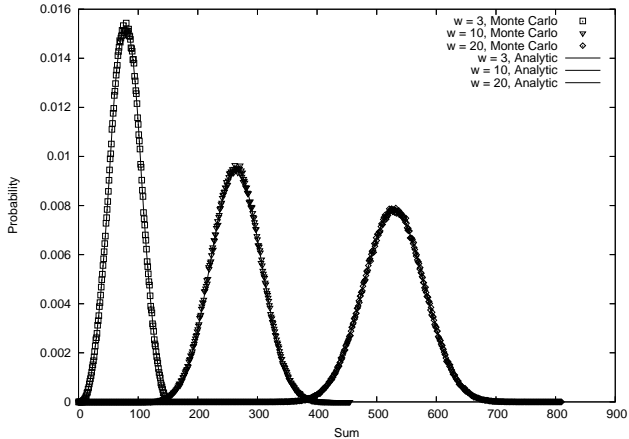}
   \caption{Probability distribution for different values of the window size $w = 3, 10, 20$, and total number of 
   points $N=50$; Monte Carlo probability distributions using $1,000,000$ samples.}
   \label{fig:PD}
\end{figure}

Figure \ref{fig:PD} shows the probability distributions for different values of $(w,N)$ 
calculated using the analytical and the Monte Carlo approach. It is important to note that $\epsilon$ and $\alpha$ are statistically motivated and must be defined. In this example, our confidence is $\alpha = 10^{-4}$ and accuracy of $10\%$ ($\epsilon = .1$). Thus, we must perform $1,000,000$ random samples. The error in the tails of the distribution seen in Figure \ref{fig:PD} is less than $10\%$, and thus consistent with theory.

%\begin{table}[t]
%\begin{center}
%\caption{Euclidean Distances for Probability Distributions shown in Figure \ref{fig:PD}\newline}
%\begin{tabular}{ | l | l | l |}
%\hline
%  $w$ & $N$ & Euclidean Distance \\
%  \hline
%  3 & 125 & $0.001022$ \\
%  4 & 125 & $0.000973$ \\
%  5 & 125 & $0.000987$ \\
%  10 & 125 & $0.000972$ \\
%  20 & 125 & $0.000963$ \\
%\hline
%\end{tabular}
%\label{table:pd_errors}
%\end{center}
%\end{table}

Not all time series are of equal length $N$, and thus we cannot use the same probability distributions for every time series. Although one could create a probability distribution for every possible $N$, this is not practical. In order to solve this problem, one can split the time series into equal parts of length $N_s$, where $N_s \leq N$ and $N_s \gg w_{max}$, where $w_{max}$ is the largest possible window size. Each subset $N_s$ is then analyzed as a single time series. To ensure that the algorithm does not miss any possible events where any two intervals meet, one should overlap the intervals by $w_{max}$. More formally, given an interval for a fixed $N$, $[v_i, v_{i+1}, \ldots, v_{i+N-1}]$, the subsequent search interval should be $[v_{i+N-w_{max}}, \ldots, v_{i+2N-w_{max}-1}]$. Furthermore, this method may be used to analyze time series of large lengths, where it is too computationally complex to compute a full probability distribution for the length of the time series. In order to ensure a correct modeling of the noise of $N$, a large $N_s$ must be chosen. This is sufficient, such that as $N$ grows large, the probability distributions become increasingly similar, and thus do not change the significance of the $p$-value.

%\begin{figure}[t] %  figure placement: here, top, bottom, or page
%   \centering
%   \includegraphics[width=3.25in,clip=true,trim = 0.8in 0.8in 0.43in 0.8in]{figures/w_3_10_20_N50_residuals.pdf}
%   \caption{Residual plot for Figure \ref{fig:PD}.}
%   \label{fig:PD_residuals}
%\end{figure}

These distributions need only be calculated once. As a benefit of having a single model based on $w, N$, one can store the distributions to be recalled for later analyses. Because all time series are first converted to \emph{rank-space} and thus have a uniform distribution of noise, we can use the same noise distribution for all time series that are of the same length.

%------------------------------------------------------------------------- 
\section{Computationally Tractable Search}
\label{optimization}

Given a probability distribution for each possible window size $w$, $1 \leq w \leq N$, and for a given $N$, we wish to find areas of significance. A na\"{i}ve approach to identifying events would be to apply a brute force technique. We compare each sum $Q(s, w)$ for all $s,w$ to the probability distribution defined by the method in Section \ref{probability_dist} or Section \ref{probability_dist2} for $w$ and $N$. The search examines the $p$-value associated with the sum $Q(s, w)$, and ranks the regions by $p$-value, lowest to highest. The complexity of the brute force search is $O(N^2)$. For very large data sets, this is unacceptable and a quicker approximation is needed. In Section \ref{surface_plot}, an analysis of the probability surface plot gathered from analyzing all possible $(s,w)$ pairs is shown to be quite revealing, and motivates our use of well-known optimization methods for approximation of results. In addition, it is important to note that we refer to \emph{event regions} rather than specific $(s,w)$ pairs, because similar pairs of $(s,w)$ may represent the same event. The algorithm outlined in Section \ref{combine_results} presents a method for combining similar (or \emph{overlapping}) windows and return representative $(s,w)$ values for each of the event intervals. 

%------------------------------------------------------------------------- 
\subsection{Examining the Probability Surface Plot.}
\label{surface_plot}

\begin{figure}[t] %  figure placement: here, top, bottom, or page
   \centering
   \includegraphics[width=3.25in,clip=true,trim = 0.8in 0.8in 0.43in 0.8in]{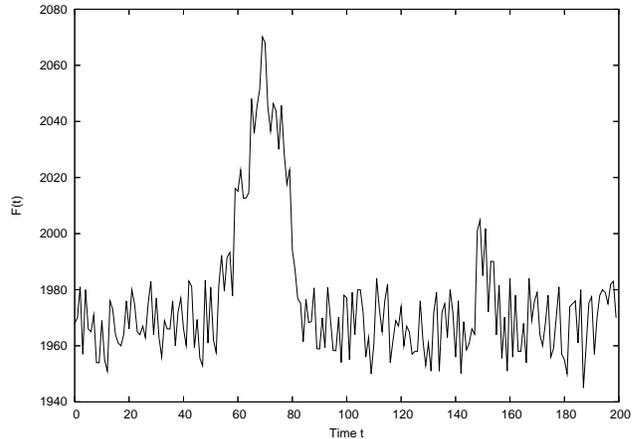}
   \caption{Synthetic time series with two events.}
   \label{fig:g_200_ts}
\end{figure}

\begin{figure}[t] %  figure placement: here, top, bottom, or page
   \centering
   \includegraphics[width=3.25in,clip=true,trim = 0.90in 2.95in 0.90in 3.1in]{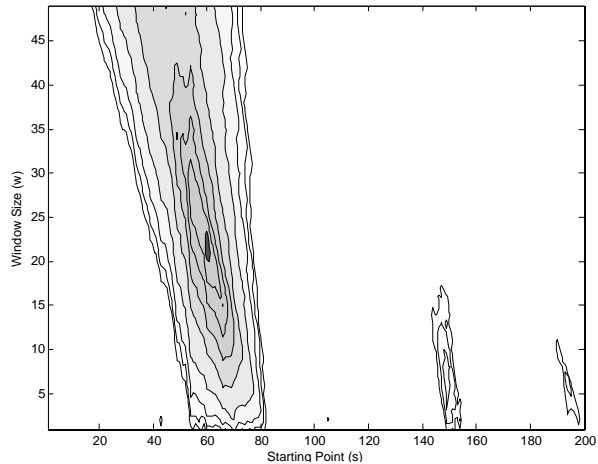}
   \caption{Contour plot of probabilities for all $(s,w)$ pairs from analysis of Figure \ref{fig:g_200_ts}.}
   \label{fig:ts_and_surface}
\end{figure}

A two-dimensional, $s \times w$, surface plot of the probabilities shows clearly where significant events occur in this space. For the time series shown in Figure  \ref{fig:g_200_ts}, Figure \ref{fig:ts_and_surface} depicts the probabilities for each pair $(s,w)$, where darker regions corresponds to lower $p$-values. After examination, one notices three regions of low-probability points. By locating the minimum of each low-probability region, one can pinpoint the exact time and duration of the event in the time series, by identifying the starting point $s$ and window size $w$. In order to find these regions algorithmically and efficiently, the problem can be reduced to the problem of minimization (a special case of the class of \emph{optimization} algorithms).

There are several methods for performing the necessary minimization \cite{simulated_annealing, simplex_method, numerical_recipes}. The success of the method detailed in this paper does not depend significantly on the particular minimization technique. Powell's method \cite{powell} was chosen due to its efficiency and the lack of need to compute derivatives. Our goal is to find all minima that are significant because we search for multiple events. Section \ref{combine_results} addresses this problem.

%------------------------------------------------------------------------- 
\subsection{Random Restarts and Combining Results.}
\label{combine_results}

Powell's method begins by selecting a search vector (a vector of points on the surface plot), performing a linear minimization on the vector, and then selecting a new search vector from this point. These steps are repeated until a satisfactory minimum is found. Random restarts are introduced to alleviate dependence on the original search vector. This helps avoid finding spurious solutions, which often appear when using approximation methods in the presence of noise. For each random restart, a pseudo-random number generator is used to seed a different vector on the surface plot. When the window size of the desired event to be found is small (i.e., when $w$ is small, the event would appear as a quick spike in the time series), any minimization method is likely to find spurious solutions. These cases would benefit from more random restarts to ensure that the significant minima are found. The analyst must define the number of random restarts required (suggested values for this parameter are discussed in Section \ref{synthetic_ts}).

\begin{figure}[h] %  figure placement: here, top, bottom, or page
   \centering
   \includegraphics[width=3.25in,clip=true,trim = 0.9in 2.75in 0.92in 2.9in]{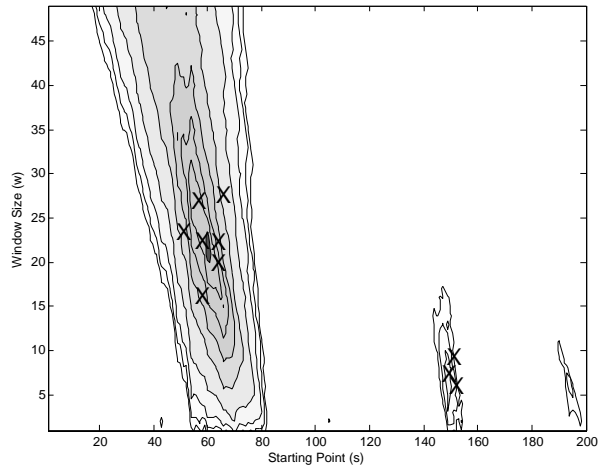}
   \caption{Contour plot of the $p$-values of the analysis of Figure \ref{fig:g_200_ts}, each X represents a minimization using Powell's method.}
   \label{fig:optimization_results}
\end{figure}

Because many random restarts will identify similar instances of the same solution (see Figure \ref{fig:optimization_results}), we introduce an agglomerative clustering method to find the best suited final $(s,w)$ pairs for an \emph{event region}. Consider two results, $(s_1, w_1)$ and $(s_2, w_2)$. The goal is to determine if they represent the same solution. Thus, we consider the \emph{overlap} of the two results, and if that overlap is within a threshold $\theta$, the two pairs are considered to be the same result and are in the same cluster $C_i$. Thus, two results overlap if
\begin{equation}
\begin{array}{l}
	\theta \leq (\rm{R_{total}} - (\rm{R_{begin}} + \rm{R_{end}})) / w_1\\
	\theta \leq (\rm{R_{total}} - (\rm{R_{begin}} + \rm{R_{end}})) / w_2
\end{array}
\end{equation}
\noindent where
\begin{equation}
\begin{array}{l}
	\rm{R_{total}} = \mathrm{max}(s_1+w_1, s_2+w_2) - \mathrm{min}(s_1, s_2)\\
	\rm{R_{begin}} = |s_1 - s_2|\\
	\rm{R_{end}} = |(s_1 + w_1) - (s_2 + w_2)|
\end{array}
\end{equation}
To build the clusters, we iterate through each result and compare the current result $(s', w')$ with each element of each cluster $C_i$. If no cluster exists in which each element overlaps with the current result, we create a new cluster $C_k$ containing only $(s', w')$. Finally, in each cluster, the element in that cluster corresponding to the sum with the lowest probability should be used as the representative $p$-value for that cluster. In experiments, we found that the results are fairly insensitive to the chosen value for $\theta$ (the default value used in our experiments was $\theta = .75$).

Note that random restarts are beneficial when dealing with time series that could have multiple events. In these situations, the random restarts will allow the minimization to find multiple intervals with low $p$-values, each of which has the potential to be an event. In the following section, we explore how to distinguish between likely events and spurious solutions when dealing with the final $(s,w)$ pairs determined from the clustering algorithm.

%\begin{overpic}[tics=5,width=3.25in]{figures/surface.pdf} % add "grid" to see grid in here (before width)
%\put(25,55){$\times$}
%\put(25,56){$\times$}
%\put(20,50){$\times$}
%\put(26,60){$\times$}
%\put(20,58){$\times$}
%\put(27,71){$\times$}
%\put(25,45){$\times$}
%%\put(32,74){\includegraphics[scale=.3]{figures/surface.pdf}}
%%\put(32,77){\llap{\scriptsize%
%%\colorbox{back}{Windm\"uhle}}}
%%\put(28,63){\small\textcolor{red}{\ding{55}}}
%%\put(6.3,13){\colorbox{back}{{\Pisymbol{ftsy}{68}}}}
%%\put(29.8,61.4){\color{blue}\vector(1,3){2}}
%%\put(38.6,63){\color{blue}\vector(1,3){2}}
%\end{overpic}
%% bit of a hack...so keep these two together...

%------------------------------------------------------------------------- 
\subsection{Identifying Events.}

After we have found all potentially significant events, we rank order them by significance for further examination. There are two aspects to consider when analyzing the \emph{event regions}. First, one should consider all intervals that are minima in the surface plot of $p$-values for a time series. Each of these minima are included in the ranking of events for all time series being considered. By identifying all minima, we are able to find multiple areas of interest in a single time series. Such is the case in Figure \ref{fig:g_200_ts}, which is an example of time series with multiple events with differing $p$-values.

\begin{figure}[t] %  figure placement: here, top, bottom, or page
   \centering
   \includegraphics[width=3.25in,clip=true,trim = 0.8in 0.8in 0.43in 0.8in]{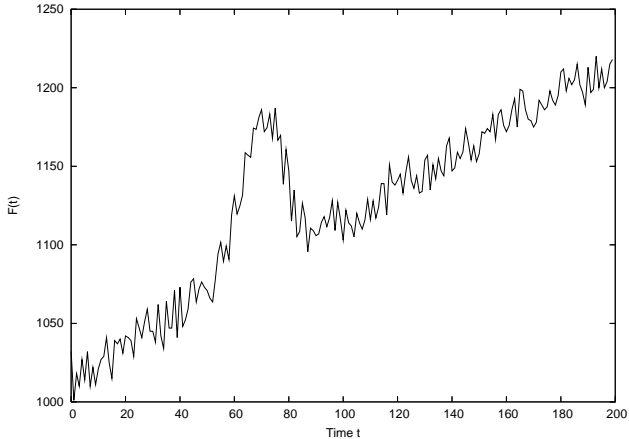}
   \caption{Synthetic time series with trend.}
   \label{fig:trend_ts}
\end{figure}

\begin{figure}[t] %  figure placement: here, top, bottom, or page
   \centering
   \includegraphics[width=3.25in,clip=true,trim = 0.9in 2.75in 0.92in 2.9in]{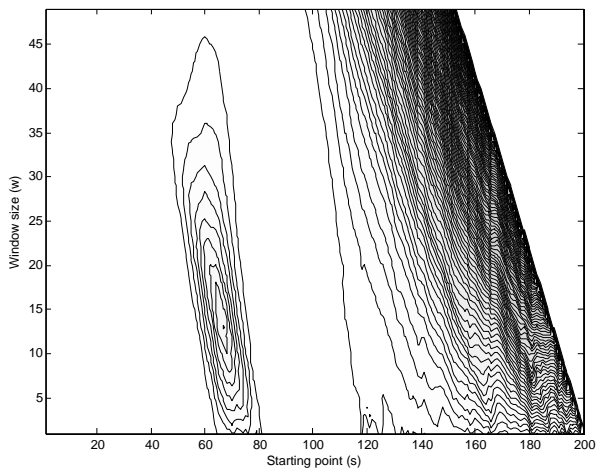}
   \caption{Contour plot of the $p$-values of the analysis of Figure \ref{fig:trend_ts}.}
   \label{fig:trend_contour}
\end{figure}

Second, in some cases, where a trend exists or the signal-to-noise ratio is low, simply ranking by the $p$-value of each true minima may result in overlooking some events. For example, Figure \ref{fig:trend_ts} is a time series with an upward trend. As one can see from the surface plot in Figure \ref{fig:trend_contour}, the most significant window resides at the end of the trend, but there is clearly another minimum \emph{with significance} at $s = 70$ and $w = 13$. Thus, it is also important to consider the frequency with which the \emph{event regions} were identified during the random restarts. This will identify the intervals that are clear minima on the probability surface plot, and thus could be significant to the time series. The frequency of the \emph{event regions} can be considered by the analyst in addition to its $p$-value.

Although our method will identify significant events on a trend like the one in Figure \ref{fig:trend_ts}, the \emph{event regions} associated with the end of the trend will still appear as significant. Thus, in the case of a data set with many trends, it may be preferable to first consider filtering the time series by detrending them. The most basic method would be to subtract the best-fit line. One can also consider multiple detrending algorithms for specific fields, such as detrending for astronomy \cite{detrend_astronomy2, detrend_astronomy1} or detrending for medical applications \cite{detrend_medical}.

%------------------------------------------------------------------------ 
\section{Results}
\label{results}

Two major analyses were performed with the method described in this paper. First, we analyze a subset of the MACHO survey that was chosen because there are known events in the data, and also because it represents a large enough size to demonstrate the method's efficiency. Our results show the accuracy of the method in our  astronomical application. Second, we create synthetic time series to find empirical results for one of the method's parameters, and in addition, show further evidence of the accuracy of the method.

%------------------------------------------------------------------------- 
\subsection{MACHO.}
\label{macho_results}

\begin{figure}[t] %  figure placement: here, top, bottom, or page
   \centering
   \includegraphics[width=3.25in,clip=true,trim = 0.8in 0.8in 0.43in 0.8in]{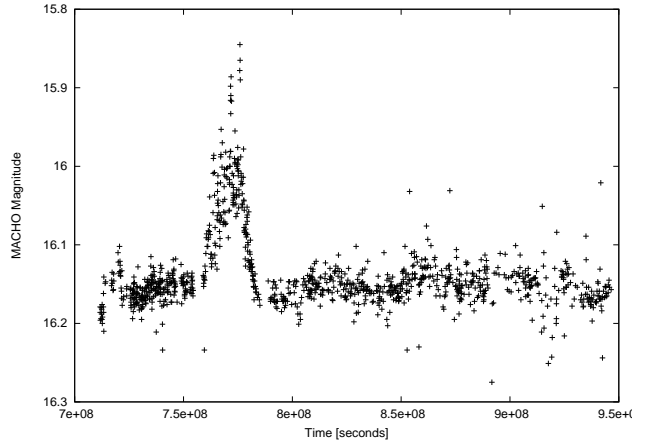}
   \caption{Blue star in MACHO survey (MACHO ID: 2.5628.4423). Event was found with $p$-value $= 5.008 \times 10^{-142}$.}
   \label{fig:bluestar_example}
\end{figure}

The MACHO project spanned several years and observed millions of stars in order to detect a few rare events. It is a well-known and well-explored survey \cite{macho_original}. Our goal is to find at least two known types of events in this data: microlensing events \cite{macho_microlensing} and evidence of blue stars \cite{macho_bluestar}. Other event types exist, but these two examples are well researched and cited, and thus serve the purpose of verifying our results. Microlensing events are characterized by a single significant deviation from the baseline (see Figure \ref{fig:event_example_ml} for an example). Blue stars are characterized by a similar shape, and also have a similar duration to a microlensing event (see Figure \ref{fig:bluestar_example} for an example) \cite{macho_bluestar}.

Our analysis was performed on $110,568$ time series from 56 tiles (from the red band), each tile representing a portion of the sky. In total, there were $28$ known microlensing events and $5$ known blue stars in the subset. Each time series had a length between $1000$ and $1812$, with an average length of $1386$. Before performing our analysis, some basic detrending was performed by subtracting a least-squares linear fit for each time series. All time series were analyzed with the same probability distribution for $N=1000$, where any time series larger would use the overlapping method described in Section \ref{probability_dist2}.

\begin{figure}[h] %  figure placement: here, top, bottom, or page
   \centering
   \includegraphics[width=3.25in,clip=true,trim = 0.8in 0.8in 0.43in 0.8in]{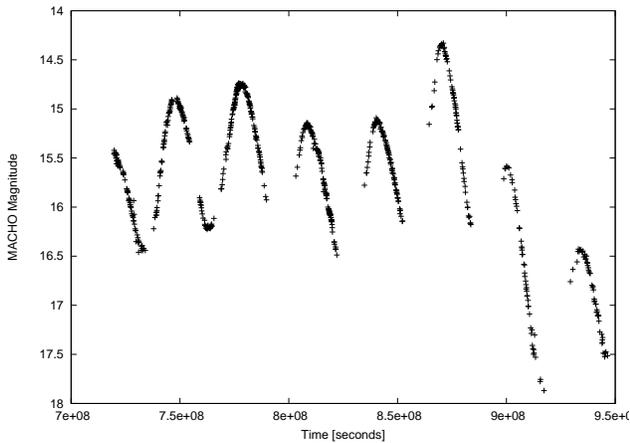}
   \caption{MACHO ID 80.6468.77. Periodic (or pseudo-periodic) event not identified as microlensing or as a blue star. $p$-value is $1.688 \times 10^{-104}$, with a rank of 55.}
   \label{fig:nonevent2}
\end{figure}

\begin{figure}[h] %  figure placement: here, top, bottom, or page
   \centering
   \includegraphics[width=3.25in,clip=true,trim = 0.8in 0.8in 0.43in 0.8in]{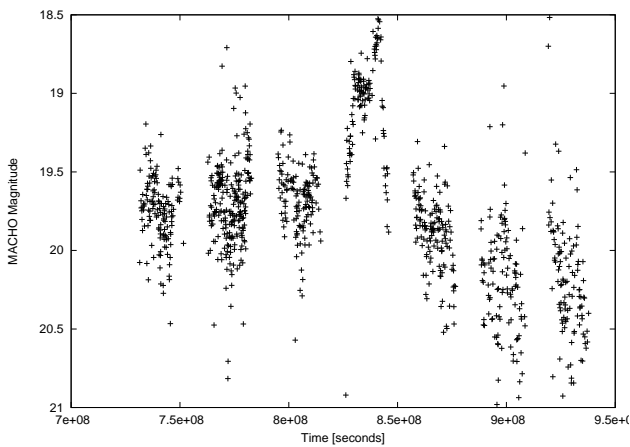}
   \caption{MACHO ID 118.18278.261. Event not identified as microlensing or as a blue star. $p$-value is $8.167 \times 10^{-112}$, with a rank of 27.}
   \label{fig:nonevent4}
\end{figure}

A ranking of the set of possible events (i.e., intervals of time series) was done by considering the event with the lowest $p$-value for each interval. Table \ref {table:macho_median} summarizes the analysis of the MACHO data set. There was no difference between the results obtained from the analytic and Monte Carlo methods. From the results, we observe that all known events were found in the top $1.1\%$ of the ranks. In order to show that this is a desirable result, one must examine the events found to be more significant than the lowest ranked known event (rank $1217$). We examined all $1217$ significant light curves and found that each either contains a significant event, or is periodic/pseudo-periodic and thus they appear as significant deviations from the baseline. Examples of an unidentified and pseudo-periodic events are found in Figures \ref{fig:nonevent2} and \ref{fig:nonevent4}, respectively. %We provide a comparison of these results with the Monte Carlo distribution found in Table \ref{table:macho_median_montecarlo}.

%\begin{figure}[h] %  figure placement: here, top, bottom, or page
%   \centering
%   \includegraphics[width=3.25in,clip=true,trim = 0.8in 0.8in 0.43in 0.8in]{figures/macho_plots/77_7307_23_0.pdf}
%   \caption{MACHO ID 77.7307.23. Event not identified as microlensing or as a blue star. $p$-value is $1.323 \times 10^{-133}$, with a rank of 10.}
%   \label{fig:nonevent1}
%\end{figure}

%\begin{figure}[h] %  figure placement: here, top, bottom, or page
%   \centering
%   \includegraphics[width=3.25in,clip=true,trim = 0.8in 0.8in 0.43in 0.8in]{figures/macho_plots/108_19595_35_1.pdf}
%   \caption{MACHO ID 108.19595.35. Event not identified as microlensing or as a blue star. $p$-value is $1.680 \times 10^{-101}$, with a rank of 59.}
%   \label{fig:nonevent3}
%\end{figure}

%[Will include info on the distribution of how many time series had how many events].

%\begin{table}[h]
%\centering
%\caption{Results of MACHO Analysis.}
%\begin{tabular}{ | l | r | r | r | }
%\hline
%  Event Type & Mean Rank & Median Rank & $p$-value of Median \\
%  \hline
%  Microlensing & $1,022$ & $365$ & $8.216 \times 10^{-34}$\\
%  Blue Star & $572$ & $113$ & $8.262 \times 10^{-60}$\\
%\hline
%\end{tabular}
%\label{table:macho}
%\end{table}

\begin{table*}[t]
\centering
\caption{Results of MACHO Analysis, analytic.}
\begin{tabular}{ | l | r | r | r | r | r | }
\hline
  Event Type & Median Rank & $p$-value of Median & Rank of Last & $p$-value of Last \\
  \hline
  Microlensing & $159$ & $6.577\times 10^{-50}$ & $1217$ & $1.102 \times 10^{-21}$ \\
  Blue Star & $114$ & $4.296 \times 10^{-61}$ & $324$ & $8.872 \times 10^{-42}$ \\
\hline
\end{tabular}
\label{table:macho_median}
\end{table*}

%\begin{table*}[t]
%\centering
%\caption{Results of MACHO Analysis, Monte Carlo.}
%\begin{tabular}{ | l | r | r | r | r | r | }
%\hline
%  Event Type & Median Rank & $p$-value of Median & Rank of Last & $p$-value of Last \\
%  \hline
%  Microlensing & $186$ & $3.951\times 10^{-45}$ & $1286$ & $2.468 \times 10^{-20}$ \\
%  Blue Star & $107$ & $9.781 \times 10^{-66}$ & $365$ & $1.350 \times 10^{-39}$ \\
%\hline
%\end{tabular}
%\label{table:macho_median_montecarlo}
%\end{table*}

\begin{figure}[h] %  figure placement: here, top, bottom, or page
   \centering
   \includegraphics[width=3.25in,clip=true,trim = 0.8in 0.8in 0.43in 0.8in]{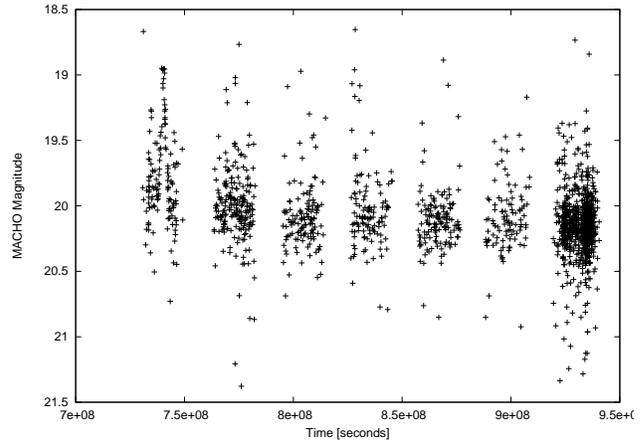}
   \caption{MACHO ID 119.20610.5200. Microlensing event with $p$-value of $4.273 \times 10^{-26}$, but with $\chi^2 = 1.13$.}
   \label{fig:low_chisq_ml}
\end{figure}

\begin{figure}[h] %  figure placement: here, top, bottom, or page
   \centering
   \includegraphics[width=3.25in,clip=true,trim = 0.8in 0.8in 0.43in 0.8in]{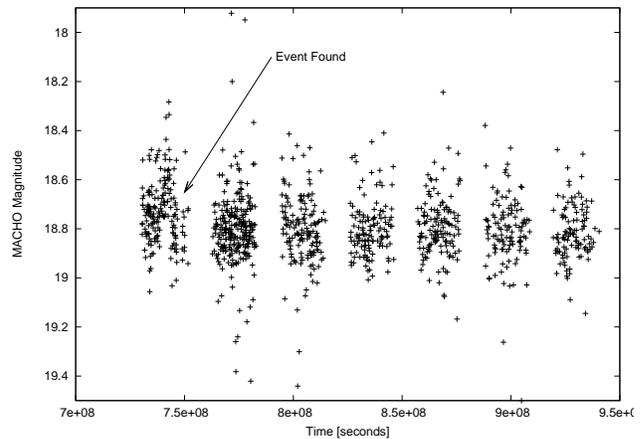}
   \caption{MACHO ID 113.18809.1561. Unidentified event with $p$-value is $3.184 \times 10^{-17}$, a rank of 1199, and $\chi^2 = 0.82$.}
   \label{fig:low_chisq_nonevent}
\end{figure}

In Section \ref{related_work}, we discuss past work for the discovery of events in astronomical data, such as performing a $\chi^2$ test. For example, the microlensing example in Figure \ref{fig:low_chisq_ml} has $\chi^2 = 1.13$, but has a $p$-value significance of $4.273 \times 10^{-26}$ and a rank in the above analysis of $689$. Moreover, it is important to note that $371$ new events with $\chi^2 < 3$ were discovered in these top ranks, such as the example in Figure \ref{fig:low_chisq_nonevent}. We are currently examining these events to determine the nature of those phenomena. Each event must be examined carefully, by comparing colors and desirably spectra information. A follow-up paper to appear in an astronomical journal will address those cases \cite{future_macho_events}.

In addition, we compared our results to HOT SAX to make the distinction between event detection and anomaly detection. The analysis was run on the same data set of $110,568$ detrended time series, and was done using the brute force method of HOT SAX. The distance of the discord was calculated using the Euclidean distance, and the results were then ranked from highest to lowest Euclidean distance (the largest discord between nearest neighbors). In the top $1217$ results, only 3 known microlensing events were discovered, with many other time series of little significance appearing as false positives. The results are summarized in Table \ref{table:hotsax_macho}. It is clear that finding significance of subintervals is not the goal of HOT SAX, which is more attuned to periodic series with less noise, where it performs quite well \cite{hotsax}. As discussed in Section \ref{related_work}, the distinction between anomaly detection and event detection is key, in that they have different goals.

\begin{table}[h]
\centering
\caption{Results of MACHO Analysis using HOT SAX.}
\begin{tabular}{ | l | r | r | r | }
\hline
  Event Type & Median Rank & Lowest Rank \\
  \hline
  Microlensing & $34,233$ & $91,779$\\
  Blue Star & $21,691$ & $52,866$\\
\hline
\end{tabular}
\label{table:hotsax_macho}
\end{table}

%------------------------------------------------------------------------- 
\subsection{Synthetic Time Series.}
\label{synthetic_ts}

In order to examine how the number of random restarts required by our method affects the results, we performed an analysis on synthetic time series. The data set consists of $2000$ synthetic time series, $323$ of which are time series with varying sizes of single events, and $50$ with two events. The remaining $1627$ time series are gaussian noise, with a standard deviation of $5.0$ and a mean of $0.0$.

For those time series with events, the events were created using the following function $f(t)$:
\begin{equation}
f(t) = h e^{\frac{-(t-S)^2}{2\theta^2}} + \epsilon
\end{equation}
\noindent where $\epsilon$ is the error (or noise), $h$ is the height of the event (Y-axis), $S$ is the time $t$ at the highest point of the event (X-axis), and $\theta$ modifies the length of the event. Our events ranged from $\theta = 1.5, \ldots, 10.5$ by increments of $0.5$, $h = 20, \ldots, 100$ by increments of $5$, and $\epsilon$ was a random variable drawn from a Gaussian distribution with a standard deviation $\sigma = 5.0$ and a mean $\mu = 0$. Thus, the signal to noise ratio of these time series ranged from $2$ to $20$.

We ran our method on all $2000$ synthetic time series, and the events were ranked by their $p$-value. All $423$ events ($323$ from single event time series, and $2$ events for each of the $50$ time series with two events) were in the top $423$ ranks, without a single false positive in those ranks. All events were found with a $p$-value of $6.018 \times 10^{-9}$ or less, whereas all of the most significant intervals in the time series without events had $p$-values higher than $2.891 \times 10^{-7}$.

HOT SAX performed almost as well on this data. In the top ranks, only $7$ false positives were reported. It is interesting to note that the lower ranked events were those that occurred in time series with two events. This is due to the fact that the algorithm will consider the two events as nearest neighbors, and subsequently each will be scored lower. As stated, the comparison of such events is ideal for situations in which a time series has periodic events and one is attempting to discover anomalies within these events, not discovering the events themselves. In addition, a second experiment was conducted with HOT SAX by adding a single point to each of the time series (at a random value between $1$ and $N$), with a value of $-5h$. Adding such a value is consistent with many domains, such as astronomy, where outliers in time series are quite common. After conducting the experiment, the results were no longer useful, with no correlation in the top ranks with time series that contained events. Our method was able to reproduce nearly identical results to the original synthetic time series experiment.

The final aspect of this experiment was to consider different values for the number of random restarts needed for our method. The above experiment was performed with $30$ random restarts. In addition, the analysis was done with $5, 10, 20, 30, 50$ random restarts. Table \ref{table:random_restarts} summarizes the results of this experiment. It is important to note that significant events in this experiment are considered those that are ranked above a $p$-value threshold of $10^{-8}$, the $p$-value for which all pure-noise time series fall below.  The column \emph{One-Event TS} represents the number of events found (one per time series) in the set of $323$ time series with only one event. The next column, \emph{Two-Event TS}, represents the number of total events found from the set of $50$ time series with two events (thus, we hope to find $100$ events). It is clear that increasing the number of random restarts also increases the possibility of finding a second significant event, but that one begins to experience diminishing returns after about $30$ random restarts. For the sake of completeness, the experiment was run with all $2000$ time series for each number of random restarts ($5, 10, 20, 30, 50$). Although no false positives were found in any of the runs, there were some false negatives (events that were not discovered) when the number of random restarts was low. It is important to note that those events not discovered when the random restarts were below $30$ are those with events consisting of very few points (where $\theta < 3$ from $f(t)$ above). This is because the \emph{event region} associated with the surface plot of the $p$-values (explored in Section \ref{surface_plot}) is quite small, and will often be missed by the optimization method.

\begin{table}[t]
\centering 
\caption{Results of synthetic time series analysis for determining the required number of random restarts}
\begin{tabular}{ | l | l | l |}
\hline
  Random Restarts & One-Event TS & Two-Event TS \\
  \hline
  5 & 93.5\% & 58\% \\ %302,58
  10 & 99.1\% & 73\% \\ %320,73
  20 & 100\% & 96\% \\ %323,96
  30 & 100\% & 100\% \\ %323,100
  50 & 100\% & 100\% \\ %323,100
\hline
\end{tabular}
\label{table:random_restarts}
\end{table}

%------------------------------------------------------------------------- 
\subsection{Timing Results.}

In this section, we present the timing of applying our method using the learned probability distributions. The analysis of a single time series includes transformation into \emph{rank-space} (Section \ref{rankspace}) and then an approximate search for results using the optimization algorithm described in Section \ref{optimization}. All $56$ tiles were analyzed in parallel on a cluster, each on a separate Xeon E5410 2.3Ghz processor with a global memory of 32,768 GB. The computation required approximately $4$ minutes $23$ seconds to complete in total, with an average of approximately $0.13$ seconds per time series. Because the probability distributions are computed only once, the method scales linearly with the number of time series, each of which computes in constant time (due to the approximation method described in Section \ref{optimization}). A discussion of the complexity of computing the initial probability distribution can be found in Section \ref{assess_significance}.

The source for this project is available at the Time Series Center website, located at \emph{http://timemachine.iic.harvard.edu/}.

%------------------------------------------------------------------------- 
\section{Conclusion}

In this work, we developed a method for identifying statistically significant events in time series without the need to model the noise, and without the restriction of a fixed-size sliding window. The current literature in event detection describes methods requiring one or both of these restrictions.

By first converting a time series to \emph{rank-space}, it allows us to build noise distribution models that generalize to any time series. All possible intervals of the new time series can then be considered, and the sums of the values inside the window can be compared to a probability distribution of possible sums to determine statistical significance. Scan statistics motivated our work in developing probability distributions for sliding windows. In addition, we showed two methods for developing these distributions, one using an analytic method that is often computationally intractable, and another using a quicker Monte Carlo method. Finally, when identifying events, we showed that minimization techniques can be used to reduce the complexity of the brute force method by approximating our results.

Our method has three parameters. First, the number of random restarts must be defined, but the sensitivity of this parameter is small, as our analysis showed that one begins to see diminishing returns after a reasonable number of random restarts. Second, if one uses the more computationally tractable method for computing the probability distribution (as outlined in Section \ref{probability_dist2}), the analyst must define $\epsilon$ (the acceptable error). Third, in Section \ref{combine_results}, the value $\theta$ was used to define what was considered the same \emph{event region}. If one chooses a $\theta$ that is too small, different events may be clustered together and one may not find all events in a time series. On the other hand, if $\theta$ is too large, the results may report the same \emph{event regions} several times. In our evaluation, the results of the method are fairly insensitive to this value, as it can range from $0.05 \leq \theta \leq 0.75$ with little change in the results.

We successfully performed our method on the MACHO data set. In a set of data full of interesting events, such as microlensing and blue stars, we were able to identify the most likely candidates to be something of interest to an astronomer. We then compared these to known events, and showed that our method was able to recover all known events. Our method identified several events that generally fail traditional tests, and furthermore, identified several events that are currently unidentified. Next, we performed the same analysis on synthetic time series of varying sizes, lengths and with different noise characteristics. The algorithm performed as expected (returning no false positives), and empirical results were presented for the only parameter required by the method, the number of random restarts for the minimization technique. Finally, to clarify the distinction between event and anomaly detection, our method was compared to a leading anomaly detection algorithm, and our method was shown to find stronger results in our motivating domain and when performing the analysis on synthetic time series.

This paper presented results from a subset of MACHO, which was large enough to understand the efficacy of our method. In the near future, we will apply our method to other large astronomical surveys such as MACHO, TAOS and OGLE. Other upcoming surveys, such as Pan-STARRS and LSST, could also benefit from using this method to analyze the millions of stars that require analysis.

%-------------------------------------------------------------------------
\section*{Acknowledgments}
We would like to thank Ira Gessel from the Mathematics Department at Brandeis University for his help.
% in helping us determine the analytic method for probability distributions outlined in Section \ref{probability_dist}.

%-------------------------------------------------------------------------
%\nocite{ex1,ex2}
\bibliographystyle{siam}

\bibliography{draft3}

\end{document}